\documentstyle[12pt,epsf]{article}
\newcommand{\be}{\begin{equation}}
\newcommand{\ee}{\end{equation}}
\newcommand{\bea}{\begin{eqnarray}}
\newcommand{\eea}{\end{eqnarray}}
\newcommand{\brr}{\begin{array}}
\newcommand{\err}{\end{array}}
\newcommand{\bc}{\begin{center}}
\newcommand{\ec}{\end{center}}
\newcommand{\nn}{\nonumber}
\newcommand{\hl}{\hline}

\newcommand{\ra}{\rightarrow}

\newcommand{\msbar}{\overline{\mbox{\scriptsize MS}}}
\newcommand{\MSbar}{\overline{\mbox{MS}}}
\newcommand{\gammaz}{\hat{\gamma}^{(0)T}}
\newcommand{\gammau}{\hat{\gamma}^{(1)T}}
\newcommand{\as}{\alpha_{ s}}

\newcommand{\DFz}{\Delta F=0}

\newcommand{\DFd}{\Delta F=2}

\newcommand{\DBu}{\Delta B=1}
\newcommand{\DBd}{\Delta B=2}

\newcommand{\BB}{B^{ 0}$--${\bar B}^{ 0}}
\newcommand{\KK}{K^{ 0}$--${\bar K}^{ 0}}

\newcommand{\J}{\hat{J}}
\newcommand{\U}{\hat{U}}

\newcommand{\W}{\hat{   W}}
\newcommand{\Z}{\hat{   Z}}

\newcommand{\lef}{(1-\gamma_5)}
\newcommand{\rig}{(1+\gamma_5)}

\def\dfrac#1#2{{\displaystyle {#1 \over #2}}}

\begin{document}
\pagestyle{empty}
\begin{flushright}
ROME1-1183/97 \\
TUM-HEP-296/97 \\
\end{flushright}

\vskip 1cm
\centerline{\LARGE{\bf{Next-to-Leading Order QCD Corrections to }}}
\centerline{\LARGE{\bf{$\DFd$ effective Hamiltonians}}}

\vskip 1cm
\centerline{\bf{ M. Ciuchini$^{a}$, E. Franco$^b$, V. Lubicz$^c$,}}
\centerline{\bf{ G. Martinelli$^{b}$, I. Scimemi$^b$ and L. 
Silvestrini$^{d}$.}}
\vskip 0.5cm
\centerline{$^a$ INFN, Sezione Sanit\`a, V.le Regina Elena 299,
00161 Roma, Italy. }
\centerline{$^b$ Dip. di Fisica,
Universit\`a degli Studi di Roma ``La Sapienza" and}
\centerline{INFN, Sezione di Roma, P.le A. Moro 2, 00185 Roma, Italy. }
\centerline{$^c$ Dip. di Fisica, Univ. di Roma Tre
and INFN, Sezione di Roma,}
\centerline{Via della Vasca Navale 84, I-00146 Roma, Italy}
\centerline{$^d$ Physik Department, Technische Universit\"at M\"unchen,}
\centerline{D-85748 Garching, Germany.}

\begin{abstract}
The most general QCD next-to-leading anomalous-dimension
matrix of all  four-fermion dimension-six $\DFd$ operators
is computed. The results of this calculation can be used in many 
phenomenological applications, among which the most important are those 
related to theoretical predictions of $\KK$ and $\BB$ mixing in several 
extensions of the Standard Model (supersymmetry, left-right symmetric 
models, multi-Higgs models, etc.), to estimates  the $B^0_s$--$\bar 
B^0_s$ width difference, and to the calculation 
of the $O(1/m_b^3)$ corrections for inclusive $b$-hadron decay rates.
\end{abstract}
\date{}
\newpage

\pagestyle{plain}
\setcounter{page}{1}
\section{Introduction}
\label{sec:intro}
Theoretical predictions of several measurable quantities, which are relevant 
in $K$-, $D$- and $B$-meson phenomenology, depend crucially on the matrix 
elements of some $\Delta F=2$ four-fermion operators. 
Examples are given by FCNC effects in SUSY extensions of the Standard 
Model~\cite{pellicani}-\cite{beautysusy2} (or other models such
as left-right symmetric ~\cite{lr} or multi-Higgs models), 
by the $B^0_s$--$\bar B^0_s$ width difference~\cite{bbd}, 
and by the $O(1/m_b^3)$ corrections in inclusive b-hadron decay rates
 (which actually depend on the matrix elements
of several four-fermion $\DFz$ operators~\cite{ns}). 
In all these cases, the relevant
operators have the form \be Q= C^{\alpha\beta\rho\sigma} ({\bar
b}_{\alpha}\Gamma q_{\beta})
  ({\bar b}_{\rho} \Gamma q_{\sigma}) \, , \label{eq:general} \ee
where $\Gamma$ is a generic Dirac matrix acting on (implicit) spinor
indices; $\alpha$--$\sigma$ are colour indices and 
$C^{\alpha\beta\rho\sigma}$ is either $\delta^{\alpha\beta}\delta
^{\rho\sigma}$ or $\delta^{\alpha\sigma}\delta
^{\rho\beta}$ (for the $1/m^3$ corrections to the inclusive
decay rates the flavour structure has the form
$(\bar b q ) (\bar q b)$).\footnote{
All the formulae of this paper refer to the $\DBd$ case. 
Their extension to generic $\DFd$ transitions is straightforward.}

All the  operators discussed in this paper  appear 
in some ``effective" theory, obtained by using the Operator
Product Expansion (OPE).
As a consequence, in all cases, three steps are necessary for obtaining 
physical amplitudes from their matrix elements:

i) matching of the original theory to the effective one
at some large energy scale; 

ii) renormalization-group evolution from the large energy scale 
to a low scale suitable for the calculation of the hadronic matrix elements 
(typically $1$--$5$~GeV);

iii) non-perturbative calculation of the hadronic matrix elements.

In this paper we present a calculation of the two-loop
anomalous dimension matrix relevant for $\Delta F=2$ 
transition amplitudes. This matrix can be used for the 
Next-to-Leading Order (NLO) renormalization-group evolution
of the Wilson coefficient functions of the effective theory
from the large to the small energy scale, step ii).
The anomalous dimension matrix includes leading and
sub-leading corrections of order $\as$ and $\as^2$. The calculation
has been performed in na\"\i ve dimensional regularization (NDR)
and we give results for different  renormalization schemes.
We have also verified that the results obtained in different schemes 
 are compatible.  We give many details on the calculation itself,
on the definition of the renormalized operators, on the relation between 
different renormalization schemes (and on the r\^ole of the 
corresponding counterterms), on the gauge invariance
of the final results etc. We also present a list containing
the contribution of all the Feynman diagrams to the anomalous 
dimension matrix.
The list may be useful to check our results and for further applications.
 In this paper we have preferred to give the results for 
the one- and two-loop anomalous-dimension matrix with as many details
as possible and postpone the phenomenological applications of the results
given here to further publications. In section~\ref{sec:admf},
the reader who is not interested
in the theoretical and technical aspects of the calculations can find
the final results for the anomalous dimension matrix of the operator basis
defined in eq.~(\ref{basef}) of subsec.~\ref{subsec:fierz}. 

Besides presenting the results for the $\DBd$ ($\DFd$) operators discussed 
here, we take the opportunity to clarify several issues related to the
regularization and renormalization dependence of the operators and of the 
corresponding Wilson coefficients. In particular, we discuss in detail the 
problems related to the precise definition of the so-called ``$\MSbar$ 
schemes", which, for composite operators, are not uniquely defined, even 
for a given regularization~\cite{Altarelli}--\cite{Ciuchini}.
We also examine the equivalence, 
and differences, of the most popular renormalization schemes, and the 
subtleties related to Regularization-Independent renormalization schemes 
(RI)~\cite{Ciuchini2}. 
 
The paper is organized as follows. In sec.~\ref{sec:gefo}, we introduce
the operators relevant for physical applications and define the operator 
basis for which the anomalous dimension matrix will be given;
a general discussion on the Wilson coefficients, their scheme-dependence
and renormalization-group evolution will be presented in sec.~\ref{sec:wc};
the strategy for the calculation of the anomalous dimension matrix in the 
$\MSbar$ and RI schemes, to be defined below, is given in sec.~\ref{sec:andim}; 
the final results, together with the one-loop matrices necessary to change
renormalization scheme, are also given for some relevant cases 
in sec.~\ref{sec:admf}.

\section{Four-fermion operators}
\label{sec:gefo}
We start this section by introducing the operators that we have in mind 
in view of future applications; we then illustrate 
the chiral and Fierz properties of the relevant operators which are used
to derive the general form of the mixing matrix.

\subsection{Operators relevant for physical applications}
\label{subsec:pa}
In this subsection, we present a list of operators which enter  the
calculation of the physical quantities mentioned in the 
introduction~\footnote{ Here and in the following we adopt the same 
notation as in the original papers.}:
\begin{itemize}

\item[1)] \underline{FCNC in SUSY extensions of the Standard Model:}

For $K$- and $B$-meson transitions, these effects have been recently analyzed
in detail in a series of papers, see for example 
refs.~\cite{pellicani}-\cite{beautysusy2}.
The relevant operators which enter the effective Hamiltonian are
\bea Q_1 &=& \bar b^\alpha \gamma_\mu \lef q^\alpha \ \bar b^\beta 
\gamma_\mu \lef 
q^\beta\, , \nn \\ 
Q_2 &=& \bar b^\alpha \lef q^\alpha \ \bar b^\beta \lef q^\beta 
\, , \nn \\ 
Q_3&=& \bar b^\alpha \lef q^\beta \ \bar b^\beta \lef  q^\alpha 
\, , \\ 
Q_4 &=& \bar b^\alpha \lef q^\alpha \ \bar b^\beta \rig q^\beta 
\, , \nn \\ 
Q_5&=& \bar b^\alpha \lef q^\beta \ \bar b^\beta \rig q^\alpha
\, , \nn \label{eq:susy} \eea 
together with the operators $\tilde Q_{1,2,3}$ which can be obtained from the operators
$Q_{1,2,3}$ by the exchange $\lef \leftrightarrow \rig$.

\item[2)] \underline{ The $B_s$--$\bar B_s$ width difference
 $\Delta \Gamma_{B_s}$:} 

At lowest order in $1/m_b$, by using the OPE, the width difference 
$\Delta \Gamma_{B_s}$ can be written in terms of  two $\DBd$ 
operators~\cite{bbd} 
\bea Q &=& \bar b \gamma_\mu \lef s \ \bar b \gamma_\mu \lef s\, , \nn
\\ 
Q_S &=& \bar b \lef s \ \bar b \lef s \, .\label{eq:dgs} \eea 
where, since the fermion bilinears are colour singlets 
($\bar b \gamma_\mu \lef s = \bar b^\alpha \gamma_\mu \lef s^\alpha$),
the colour indices have not been shown explicitly.

\item[3)] \underline{Heavy-hadrons lifetimes ($\tau_B$, $\tau_{B_s}$,
$\tau_{\Lambda_b}$):} 

In this case, the $1/m_b^3$ corrections to the lifetime, due to Pauli 
interference and $W$-exchange, can be written in terms of four 
operators \cite{ns} 
\bea\label{eq:lt} O^q_{V-A} &=& \bar b \gamma_\mu \lef q 
\ \bar q \gamma^\mu \lef b \, , \nn \\ 
O^q_{S-P} &=& \bar b \lef q \ \bar q \rig  b\, , \\ 
T^q_{V-A} &=& \bar b t^A \gamma_\mu \lef q \ \bar q t^A \gamma^\mu 
\lef b \, , \nn \\ 
T^q_{S-P} &=& \bar b t^A \lef q \ \bar q t^A \rig b \nn \, .
\nn \eea 
where an implicit sum over colour indices is understood. The operators 
above are $\Delta B=0$ operators. They contribute to the decay rates of 
the $B$-mesons (and $\Lambda_b$s) not only through the so-called 
``eight" diagrams, but also through tadpole diagrams, in which the light- 
or heavy-quark fields are contracted in a loop. These ``non-spectator" 
diagrams mix the operators of the basis (\ref{eq:lt}) with the lower 
dimension operators $\bar b b$, $\bar b\vec D^2 b$, $\bar b 
\sigma_{\mu\nu} G^{\mu \nu} b$, $\bar q q $, etc. The mixing matrix is, 
however, triangular. Thus, it is possible to compute separately, at the 
NLO, the $4 \times 4$ sub-matrix related to the mixing of the operators 
appearing in (\ref{eq:lt}) among themselves. For this sub-matrix, the 
Feynman diagrams entering  the calculation are the same as those 
relevant for the $\Delta B=2$ operators.
\end{itemize}

\subsection{Chiral and Fierz properties of the operators}
\label{subsec:opba}
The operators considered in 1)-2) can be expressed in terms of linear 
combinations of independent operators, defined by their colour-Dirac structure,
belonging to some basis.  In case 3), for the sub-matrix
considered here, the same colour-Dirac structure (with obvious replacement
of the flavour indices) can also be used. The choice of the basis of reference
is, however, arbitrary, and different equivalent possibilities exist.
We first present one possible choice, which we find particularly convenient
to discuss the chiral properties of the operators:
\bea
\label{base}
Q_{V_LV_L} & = & \bar \psi_1^\alpha \gamma_\mu \lef \psi_2^\alpha \:
       \bar \psi_3^\beta \gamma_\mu \lef \psi_4^\beta \nn \\
Q_{V_LV_R} & = & \bar \psi_1^\alpha \gamma_\mu \lef \psi_2^\alpha \:
       \bar \psi_3^\beta \gamma_\mu \rig \psi_4^\beta \nn \\
Q_{RL} & = & \bar \psi_1^\alpha \rig \psi_2^\alpha \:
      \bar \psi_3^\beta \lef \psi_4^\beta \nn \\
Q_{LL} & = & \bar \psi_1^\alpha \lef \psi_2^\alpha \:
      \bar \psi_3^\beta \lef \psi_4^\beta \nn \\
Q_{T_LT_L} & = & \bar \psi_1^\alpha \sigma_{\mu\nu} \lef \psi_2^\alpha \:
       \bar \psi_3^\beta \sigma_{\mu\nu} \lef \psi_4^\alpha \nn \\
\tilde{Q}_{V_LV_L} & = & \bar \psi_1^\alpha \gamma_\mu \lef \psi_2^\beta \:
       \bar \psi_3^\beta \gamma_\mu \lef \psi_4^\alpha \\
\tilde{Q}_{V_LV_R} & = & \bar \psi_1^\alpha \gamma_\mu \lef \psi_2^\beta \:
       \bar \psi_3^\beta \gamma_\mu \rig \psi_4^\alpha \nn \\
\tilde{Q}_{RL} & = & \bar \psi_1^\alpha \rig \psi_2^\beta \:
      \bar \psi_3^\beta \lef \psi_4^\alpha \nn \\
\tilde{Q}_{LL} & = & \bar \psi_1^\alpha \lef \psi_2^\beta \:
      \bar \psi_3^\beta \lef \psi_4^\alpha \nn \\
\tilde{Q}_{T_LT_L} & = & \bar \psi_1^\alpha \sigma_{\mu\nu} \lef \psi_2^\beta \:
       \bar \psi_3^\beta \sigma_{\mu\nu} \lef \psi_4^\alpha \, ,
       \nn
\eea
where $\sigma_{\mu\nu}\equiv 1/2 [\gamma_\mu,\gamma_\nu]$. 
In (\ref{base}), the flavours $\psi_1$--$\psi_4$ are all different and
the operators belong to irreducible representations of the chiral
group. To these
10 operators, we have to add those which can be obtained by exchanging
left- with right-handed fields. Since, however, strong interactions cannot
change chirality~\footnote{ In mass-independent 
renormalization schemes, such as the RI schemes discussed in 
this paper, chiral-symmetry relations, which can be derived in the massless 
theory, remain true also in the massive case.}, the second set of operators
does not mix with the operators defined in (\ref{base}) and, because
parity is conserved, the Anomalous Dimension Matrix (ADM) is the same
in the two cases. Thus, in the following, we will only consider the operators
of eq.~(\ref{base}).

The previous considerations hold  only if one uses a renormalization
prescription which preserves chirality (in this respect parity is never a 
serious issue). It often happens, e.g. in dimensional schemes such as 
the t'Hooft-Veltman $\MSbar$ one (HV), that the renormalization procedure 
violates  either chirality, or (and) other symmetries that are manifest at 
the tree level, for example the Fierz transformation properties. In order 
to simplify the presentation of the results, we will use in the following 
a renormalization scheme which preserves all the relevant symmetries 
(chirality and Fierz). With such a choice, it is then sufficient to consider 
the basis (\ref{base}). This renormalization scheme has been recently 
called the Regularization Independent (RI) scheme~\cite{Ciuchini2} 
(MOM in the early literature) to emphasize that the renormalization
conditions are independent of the regularization, although they depend 
on the external states used in the renormalization procedure and on the
gauge. The RI scheme offers also a great computational advantage in the 
calculation of the counterterms which contribute at the two-loop level: 
as demonstrated in subsec.~\ref{subsec:eotladm}, in this scheme it is 
not necessary to identify and subtract separately the counterterms relative 
to the mixing with the ``Effervescent Operators" (EOs) which appear in
 dimensional regularization~\cite{Altarelli}--\cite{Ciuchini}. 
The relation 
between the operators renormalized in some RI scheme and, for instance,
 those of the standard
$\MSbar$ schemes can then be easily found with a simple one-loop calculation.
Finally, the RI scheme allows to use, without any further perturbative
calculation, the matrix elements of the operators computed in lattice 
simulations and renormalized non-perturbatively~\cite{NP}--\cite{JAPBK}. 

Chiral symmetry, and Fierz rearrangement, have further consequences, since 
they forbid the mixing between some of the operators appearing in 
eq.~(\ref{base}). 
For this reason the ADM, $\hat \gamma$, is a block-matrix which only 
allows mixing between sub-sets of the possible operators.
In the convenient representation in which the
operators appearing in~(\ref{base}) are components of row vectors
\begin{itemize} \item[I)] $\vec Q_I \equiv (Q_{V_LV_L},\tilde{Q}_{V_LV_L})$,
\item[II)] $\vec Q_{II} \equiv (Q_{V_LV_R}, \tilde{Q}_{V_LV_R})$,
\item[III)] $\vec Q_{III} \equiv (Q_{RL}, \tilde{Q}_{RL})$,
\item[IV)] $\vec Q_{IV}
\equiv (Q_{LL},\tilde{Q}_{LL},Q_{T_LT_L},\tilde{Q}_{T_LT_L})$,
\end{itemize}
Fierz rearrangement imposes the following restrictions on the form of the mixing
matrix:
\begin{itemize}
\item For set I) the structure is given by
\be
\hat \gamma_I \equiv \pmatrix{A_I & B_I \cr B_I & A_I \cr}
\, ;\ee
\item If the mixing-matrix for II) is given by
\be
\hat \gamma_{II} \equiv \pmatrix{A_{II} & B_{II} \cr C_{II} & D_{II} \cr}
\, ,\ee
then we have 
\be
\hat \gamma_{III} \equiv \pmatrix{D_{II} & C_{II} \cr B_{II} & A_{II} \cr}
\, .\ee
\item In order to discuss Fierz rearrangement for the sector IV), we introduce
the Fierz transformation matrix for the sub-basis $\vec Q_{IV}$
 \be
{\cal F}=\pmatrix{ 0 & -\frac{1}{2} & 0 & \frac{1}{8} \cr
     -\frac{1}{2} & 0 & \frac{1}{8} & 0 \cr
     0 & 6 & 0 & \frac{1}{2} \cr
     6 & 0 & \frac{1}{2} & 0 \cr}
\, . \ee
The anomalous dimension matrix must satisfy the relation
\be
\hat \gamma_{IV}= {\cal F} \, \hat \gamma_{IV} \, {\cal F}
\ee
The mixing matrix has a simple form in the Dirac-Fierz basis
\bea
\label{base2} \vec Q_F &\equiv& (Q_1^F,Q_2^F,\tilde{Q}_1^F,\tilde{Q}_2^F)\, ,
\,\,\,\,\,  \nn \\
Q_1^F &=& Q_{LL}+\frac{1}{4}Q_{T_LT_L}\, , \nn \\
Q_2^F &=& Q_{LL}-\frac{1}{12}Q_{T_LT_L} \, , \\
\tilde{Q}_1^F &=& \tilde{Q}_{LL}+\frac{1}{4}\tilde{Q}_{T_LT_L} \, , \nn\\
\tilde{Q}_2^F &=& \tilde{Q}_{LL}-\frac{1}{12}\tilde{Q}_{T_LT_L}\, . \nn
\eea
In this basis, the ADM can be written as\be
\hat\gamma_{IV} \equiv \pmatrix{A_{IV} & B_{IV} & C_{IV} & D_{IV} \cr 
E_{IV} & F_{IV} & G_{IV} & H_{IV} \cr
     C_{IV} & -D_{IV} & A_{IV} & -B_{IV} \cr -G_{IV} & H_{IV} & -E_{IV}
 & F_{IV} \cr}
\label{fierzcond}
\ee
\end{itemize}
In summary, we have seen that the ADM for all the operators appearing 
in~(\ref{base}) can be expressed in terms of 14 quantities, i.e.
$A_I$, $B_I$, $A_{II}$, $\dots$, $H_{IV}$.

\subsection{The Fierz basis} 
\label{subsec:fierz}
In the case in which $\psi_1=\psi_3$ (or $\psi_2=\psi_4$), not all
the operators in (\ref{base}) are independent. In order
to take into account the simplifications occurring in
this particular case, it is more convenient to give the results in the 
(Fierz) basis
\bea \label{basef}
\vec Q^{\pm} &\equiv& (Q^{\pm}_1,Q^{\pm}_2,Q^{\pm}_3,Q^{\pm}_4,Q^{\pm}_5)
\nn \\ 
Q^{\pm}_1 & = &\frac{1}{2}
    \left( \bar \psi_1^\alpha \gamma_\mu \lef \psi_2^\alpha \:
       \bar \psi_3^\beta \gamma_\mu \lef \psi_4^\beta 
\pm ( \psi_2 \leftrightarrow \psi_4 )\right) \nn \\
Q^{\pm}_2 & = & \frac{1}{2}
    \left( \bar \psi_1^\alpha \gamma_\mu \lef \psi_2^\alpha \:
       \bar \psi_3^\beta \gamma_\mu \rig \psi_4^\beta 
\pm ( \psi_2 \leftrightarrow \psi_4 ) \right) \\
Q^{\pm}_3 & = & \frac{1}{2}
    \left( \bar \psi_1^\alpha \rig \psi_2^\alpha \:
      \bar \psi_3^\beta \lef \psi_4^\beta 
\pm ( \psi_2 \leftrightarrow \psi_4 )\right) \nn \\
Q^{\pm}_4 & = & \frac{1}{2}
    \left( \bar \psi_1^\alpha \lef \psi_2^\alpha \:
      \bar \psi_3^\beta \lef \psi_4^\beta 
\pm ( \psi_2 \leftrightarrow \psi_4 ) \right)\nn \\
Q^{\pm}_5 & = & \frac{1}{2}
    \left( \bar \psi_1^\alpha \sigma_{\mu\nu} \lef \psi_2^\alpha \:
       \bar \psi_3^\beta \sigma_{\mu\nu} \lef \psi_4^\beta 
\pm ( \psi_2 \leftrightarrow \psi_4 )\right) \, . \nn 
\eea
In this case, the operators do not belong, in general,
 to irreducible representations of
the chiral group, e.g. both right- and left-handed $\psi_2$ fields
appear in $Q^{\pm}_3$. 
The $\DBd$ operators are obtained from the $Q_i^+$s, by taking 
 $\psi_1=\psi_3=b$ and $\psi_2=\psi_4=q$ (with this choice of 
 flavours the $Q_i^-$ vanish). 
 In the basis (\ref{basef}), the ADM has the form
\be \hat 
\gamma^{\pm} \equiv \pmatrix{
A^{\pm} & 0 & 0 & 0 & 0\cr
0 & B & \pm C & 0 & 0\cr
0 & \pm D & E & 0 & 0\cr
0 & 0 & 0 & F^{\pm} & G^{\pm} \cr
0 & 0 & 0 & H^{\pm} & I^{\pm} \cr} \, , 
\label{str1}
\ee
and there is no mixing between the $Q^+_i$ and the $Q^-_i$ operators.

The correspondence between the operators of the basis~(\ref{basef})
and the operators which are relevant for the physical applications
listed in subsec.~\ref{subsec:pa} is the following

\bea &1)& \quad Q_1 \to Q_1^+ \, , \,\,\,\, Q_2 \to Q_4^+ 
\, , \,\,\,\, Q_3 \to -\frac{1}{2} (Q^+_4-\frac{1}{4}Q^+_5) \, , \nn \\
& & Q_4 \to Q_3^+ \, , \,\,\,\, Q_5 \to -\frac{1}{2} Q^+_2 \, ; \\
 &2)& \quad  Q \to Q^+_1 \, , \,\,\,\, Q_S \to Q_4^+ \, ; \\
  &3)& \quad O^q_{V-A} \to Q^+_1+Q^-_1 \, , \,\,\,\, Q^q_{S-P} \to Q_3^+ +Q_3^-
\, , \,\,\,\, \nn \\
& &T^q_{V-A} \to \frac{1}{2} \left(1 - \frac{1}{N_c} \right) Q^+_1 
-\frac{1}{2} \left(1 + \frac{1}{N_c} \right) Q^-_1 \\
& & T^q_{S-P} \to -\frac{1}{2N_c} \left(
Q_3^+ + Q_3^- \right) - \frac{1}{4}\left( Q_2^+ - Q_2^- \right) \, . \nn \eea

\section{The Wilson coefficients}
\label{sec:wc}
The general method  for the calculation of the Wilson coefficients,
and a detailed discussions on their renormalization-scheme dependence, 
can be found in the literature~\cite{Ciuchini2}--\cite{Ciuchini}.
In this section, we only summarize the main formulae which are necessary 
to present our results. We also take the opportunity to clarify some 
important subtleties about the renormalization-scheme dependence.

\subsection{Effective theories and Wilson coefficients}
\label{subsec:et}

In all cases of interest, the matrix elements of the effective Hamiltonian 
can be written as
\be 
\langle F \vert {\cal H}_{eff}\vert I \rangle = \sum_{i}
\langle F \vert Q_{ i}(\mu) \vert I \rangle C_{ i}(\mu) 
\, , \label{wope} 
\ee
where the $Q_i(\mu)$s are the relevant operators renormalized at the 
scale $\mu$ and the $C_i(\mu)$s are the corresponding Wilson coefficients.
We represent the operators as row vectors $\vec Q$, as in 
subsec.~\ref{subsec:opba}, and the coefficients, $\vec C(\mu)$, as 
column ones. The vectors $\vec C(\mu)$ are
expressed in terms of their counter-part,
computed at a large scale $M$, through 
the renormalization-group evolution matrix $\W[\mu,M]$
\be \vec C(\mu) = \W[\mu,M] \vec C(M)\, . \label{evo} \ee
The initial conditions for the evolution equations,
$\vec C(M)$, are obtained by matching the full theory, which includes
propagating heavy-vector bosons ($W$ and $Z^0$), 
the top quark, SUSY particles, etc.,
to the effective theory where the $W$,
$Z^0$, the top quark and all the heavy particles
have been removed simultaneously. In general,
$\vec C(M)$ depend on the definition of the operators
in a given renormalization scheme. 
The coefficients $\vec C(\mu)$ obey the renormalization-group equations:
\be \Bigl[ - \frac {\partial} {\partial t} + \beta ( \as )
\frac {\partial} {\partial \as} + \beta_\lambda( \as )
\lambda\frac {\partial} {\partial \lambda} - 
\frac {\hat \gamma^T ( \as) }{2} \Bigr] 
\vec C(t, \as(t), \lambda(t)) =0 \, , \label{rge} \ee
where $t=\ln ( M^2 / \mu^2 )$.  The term proportional to $\beta_\lambda$,
the $\beta$-function of the gauge parameter $\lambda(t)$ (for covariant gauges),
takes into account the gauge dependence of the Wilson coefficients 
in gauge-dependent renormalization schemes, such as the 
RI scheme~\cite{Ciuchini2,Altarelli}~\footnote{ In the following,
we will denote by $\lambda=1$ the Feynman gauge
and $\lambda=0$ the Landau gauge.}. This term,
the r\^ole of which 
will be discussed extensively in sec.~\ref{sec:andim},  is absent in standard 
 $\MSbar$ schemes, independently of the regularization 
which is adopted (NDR, HV or DRED for example)~\cite{bw}--\cite{Ciuchini}. 
The factor of $2$ in eq.~(\ref{rge}) normalizes the anomalous dimension
matrix as in refs.~\cite{Ciuchini2}-\cite{Ciuchini}. To simplify the 
discussion, we only consider the case where there is no crossing of 
a quark threshold when going from $M$ to $\mu$. The relevant formulae 
for the general case can be found in refs.~\cite{bjl}--\cite{Ciuchini}.

At the next-to-leading order, we can write
\bea
\W[\mu,M]  = \hat M[\mu] \U[\mu, M] \hat M^{-1}[M] \, , 
 \label{monster} \eea
where $\U$ is the leading-order evolution matrix
\be
\U[\mu,M]=  \left[\frac{\as (M)}{\as (\mu)}\right]^{
      \gammaz / 2\beta_{ 0}} \, ,
\label{u0} \ee
and the NLO matrix is given by
\be \hat M[\mu] =
 \hat 1 +\frac{\as (\mu)}{4\pi}\J[\lambda(\mu)] \, .
\label{mo2} \ee

By substituting the expression of the $\vec C(\mu)$  given in 
eq.~(\ref{evo}) in the re\-nor\-ma\-li\-za\-tion-group equations (\ref{rge}), 
and using $\hat W[\mu, M]$ written as in eqs.~(\ref{monster})--(\ref{mo2}),
we find that the matrix $\J$ satisfies the equation
\be
\J+\frac{\beta^0_\lambda}{\beta_0}
\lambda \frac{\partial \J}{\partial \lambda}
- \left[\J,\frac{\gammaz}{2\beta_{ 0}}\right] =
     \frac{\beta_{ 1}}{2\beta^2_{ 0}}\gammaz-
     \frac{\gammau}{2\beta_{ 0}} \, . \label{jj}
\ee
In eqs. (\ref{u0}) and (\ref{jj}), $\beta_{ 0}$,  $\beta_{ 1}$
and $\beta^0_\lambda$ are the first coefficients of the 
$\beta$-functions of $\as$ and of $\lambda$, respectively; 
$\hat \gamma^{(0)}$ and $\hat \gamma^{(1)}$ are the LO and NLO anomalous 
dimension matrices to be defined in sec.~\ref{sec:andim}.
$\hat U$ is determined by the LO anomalous
dimension matrix $\hat \gamma^{(0)}$ and 
is therefore regularization and renormalization-scheme independent; 
 at this order, $\lambda \partial \J / \partial \lambda$
is also regularization (but not renormalization) scheme independent;
the two-loop anomalous dimension matrix $\hat \gamma^{(1)}$, 
and consequently $\J$ and $\hat W[\mu , M]$, are, instead, 
renormalization-scheme dependent.

\subsection{Coefficient functions and scheme dependence}
\label{subsec:cf}
In this subsection, we recall some basic aspects of the calculation of the
Wilson coefficients and discuss in detail the issues of the regularization
and renormalization dependence of the coefficients and of the corresponding 
operators. We believe that this discussion may be  useful to clarify some
misunderstandings that can be found in the literature. 

In order to compute the Wilson coefficients at
a large energy scale 
$\mu \sim M$, we should consider the full set of
current-current, box and penguin
diagrams in the full theory, i.e. with propagating
heavy particles, including the $O(\as)$ corrections.
To date, for the $\DFd$ transitions, this part of the calculation
has been carried out only  in the Standard
Model and 2HDM cases~\cite{bjw}--\cite{urban}. \par 
In the full theory, the direct calculation of the current-current, box and
penguin diagrams, including $O(\as)$ corrections has the form
\be \langle {\cal H}_{eff} \rangle \sim 
\langle \vec Q^{(0) \, T} \rangle \cdot \Bigl[ \vec T^{(0)} +
\frac {\as} { 4 \pi} \vec T^{(1)}
\Bigr] = \langle \vec Q^T (\mu) \rangle \cdot \vec C(\mu) \, , \label{coe1} \ee
where $\langle \vec Q^{(0) \, T} \rangle $ are the tree-level matrix 
elements and the vector $\vec T^{(1)}$ depends
 on the external quark (and gluon) states chosen for the calculation. 
By inserting the renormalized operators of the effective
Hamiltonian, we then compute, at order $\as$, the
one-loop diagrams between the same external states as in the full 
theory, using the same regularization. In this case we 
obtain~\footnote{ The most convenient method to define the matrix elements
is by projectors on the tree-level colour-Dirac structures of the operators
belonging to the four-dimensional basis~\cite{DS=2}.}
\be \langle \vec Q (\mu) \rangle = \Bigl( 1 +
\frac {\as} { 4 \pi} \hat r
\Bigr) \langle \vec Q^{(0)} \rangle\, . \label{coe2} \ee 
The coefficients
$\vec C(\mu)$ are obtained by comparing eq.~(\ref{coe1})
with eq.~(\ref{coe2}); if we (formally) choose the renormalization scale
$\mu = M$, all the logarithms related to 
anomalous dimensions of the operators disappear and 
\be \vec C(M) = \vec T^{(0)} + \frac {\as} { 4 \pi}
\Bigl( \vec T^{(1)} - \hat r^T \vec T^{(0)} \Bigr) \, .\label{coe3} \ee
 $\vec T^{(1)}$ and $\hat r^T$ 
 depend on the external states. However their
difference depends only on the renormalization scheme, but not on the
external states. For this reason, the dependence 
on the external momenta ($\sim \ln (-p^2)$)   of $\hat r$ in  $\vec C(M)$
and $\vec T^{(1)}$ cancels out,   for details see
refs.~\cite{bjl}--\cite{Ciuchini}. In the following $\hat r$ 
and $\vec T^{(1)}$ denote  only  non-logarithmic terms.

For given external states, and for a given gauge, the matrix $\hat r$
completely specifies the renormalization scheme ($\MSbar$, RI, etc.).
In this respect, all the renormalization schemes,
including the $\MSbar$ ones, are {\it regularization
independent}. The $\MSbar$ schemes simply amount to some specific choice of
$\hat r$. This is also demonstrated by the following observation:
even when the regularization is specified, for example
the NDR one, the so-called ``$\MSbar$ scheme" is not unique. 
The renormalized operators, and consequently $\hat r$,
 depend in general on the basis chosen 
in the regularized theory to implement the minimal subtraction procedure, 
the projectors, i.e. the definition of the EOs,
etc.~\cite{bjl}--\cite{Ciuchini}. Thus, in order to define completely the
 ``$\MSbar$ scheme", we should specify all the variables (regularized basis, 
EOs, etc.) entering  the calculation. In practice, this is equivalent
to fix $\hat r$, i.e. the renormalization prescription.
 Summarizing, the {\it regularization dependence} must always be 
 understood as a ``renormalization-scheme dependence". 
 In all the
renormalization schemes (both $\MSbar$ and RI), the same information is 
contained in $\hat r$, once that the external states and the gauge are 
specified: we will then use
 eq.~(\ref{coe2}) to define the renormalized operators. The explicit expressions
of the matrix $\hat r$ in the different schemes (and the corresponding states
and gauge) can be found in sec.~\ref{sec:andim}. 

In subsection~\ref{subsec:eotladm}, it will be shown that the combination
\be \hat G = \hat \gamma^{(1)} - \Bigl[ \hat r , \hat \gamma^
{(0)} \Bigr] - 2 \beta \hat r -2\beta
_\lambda ^0\lambda \frac{\partial \hat r}{\partial \lambda }\label{hr} \ee
is renormalization-scheme independent. It can be easily shown
that a consequence of eq.~(\ref{hr}) is the independence of
the combination
\be \hat J_{RI} = \J + \hat r^T \label{eq:jri} \ee of 
the renormalization scheme (but not of the external states and, in 
general, of the gauge on 
which $\hat r$ is computed), see also refs.~\cite{Ciuchini2}--\cite{Ciuchini}.
\par
The independence of $\J_{RI}$ (and of $\hat G$)
of the renormalization scheme means the following.
As mentioned above, for given external states, and in a given gauge,
we compute the matrix element of the {\it renormalized} operators.
These operators are renormalized in some scheme, for example one of the
possible $\MSbar$ schemes.
If we change scheme, $\hat r$ and $\J$ will accordingly change, whilst
$\J_{RI}$ will remain the same. Using eqs.~(\ref{jj}) and (\ref{hr}),
we find indeed 
\be\label{jj1}
\J_{RI}+\frac{\beta^0_\lambda}{\beta_0}
\lambda \frac{\partial \J_{RI}}{\partial \lambda}
- \left[\J_{RI},\frac{\gammaz}{2\beta_{ 0}}\right] =
     \frac{\beta_{ 1}}{2\beta^2_{ 0}}\gammaz-
     \frac{\hat G^T}{2\beta_{ 0}} \, . 
\ee
The scheme independence of $\hat G$ and $\hat \gamma^{(0)}$ guarantees
the independence of the solution of eq.~(\ref{jj1}). In turn, this implies 
the renormalization-scheme independence of the matrix $\J_{RI}$.
Note also that, since $\J_{\msbar}$ is gauge independent, $\partial 
\J_{RI}/\partial \lambda= \partial \hat r^T_{\msbar}/\partial 
\lambda$. 

The renormalization-invariant properties discussed above can be used
to introduce schemes which respect all the symmetries of the 
tree-level theory.
Using eqs.~(\ref{monster})--(\ref{mo2}) and (\ref{coe3}), we introduce
a new set of Wilson coefficients $ {\vec C}'(M)$
\be \vec C(\mu) = \hat M[\mu]\, \hat U[\mu,M]\, \hat N^{-1}[M]
\, {\vec C}'(M) \, , \label{cc} \ee
where $\hat M[\mu]$ has been defined in eq.~(\ref{mo2}) and
\be \hat N[M]=
 \hat 1 + \frac{\as (M)}{4\pi}
\left( \J + \hat r^T  \right) \, , \,\,\,\,\,\,\,
\vec C'(M) = \vec T^{(0)} + \frac {\as} { 4 \pi}
 \vec T^{(1)} \, . \label{rir1} \ee
In the above equations we have neglected higher order terms
in $\as$. 
 By a suitable change of the renormalization scheme, corresponding to 
\be {\vec V}^T(\mu)= \vec Q^T(\mu) \left( 1 - \frac {\as
(\mu)} {4 \pi} \hat r^T \right) \, ,
\ee
and $ \hat M[\mu] \rightarrow \hat N[\mu] $
one gets 
\be \label{cpri} 
\vec C'(\mu) = {\hat N}[\mu] \hat U[\mu, M]
\hat N^{-1}[M] {\vec C}'(M) \, .
\ee
Equation~(\ref{cpri}) has the following interpretation: 
it corresponds to the general expression (\ref{monster}), with
the matrices $\hat M[\mu]$ and $\hat M[M]$ given in terms of $\J_{RI}$, 
which satisfies eq.~(\ref{jj1}). In this renormalization scheme,
 $\hat G = \gamma^{(1)}_{RI}$, since  $\hat r = \hat r_{RI}=0$.
Clearly, the result is independent of the renormalization scheme of the
original operators $\vec Q(\mu)$. The scheme dependence is implicitly
contained in the states and the gauge on which $\hat r$ is computed.
Thus, in the following, we will call FRI (LRI) the scheme with
matrix elements computed in the Feynman (Landau) gauge.

A remark is in order at this point. 
In refs.~\cite{desh}--\cite{paver}, for $\DBu$ transitions, the authors used 
the so-called regularization-scheme-independent coefficients (corresponding 
to our $C'(\mu)$) introduced in ref.~\cite{bjl}~\footnote{Indeed the 
renormalization scheme of ref.~\cite{bjl} has never been completely specified, 
because the external states, on which the renormalization conditions were 
imposed, have not been given explicitly.} and computed in ref.~\cite{desh}. 
In this reference, the preference for using this particular renormalization 
scheme was justified with the argument that 
also the operator matrix elements computed with factorization 
(and used in \cite{desh}) are scheme independent. It was therefore
argued that the regularization-independent coefficients are more suited
to obtain the physical amplitudes. This argument is clearly illusory:
the coefficients, though regularization independent, depend
on the external states and on the gauge at which the 
renormalization conditions have been imposed. There is no way
to match the external quark and gluon states, used in the perturbative
calculation of the Wilson coefficients, to the hadronic states
on which the operator matrix elements are computed (not to speak about gauge
invariance). A similar argument applies to the scheme used in ref.~\cite{ag},
where the authors try to get rid of the $\mu$ dependence of the 
non-leptonic amplitudes computed with factorization.
In any case, in the absence of a consistent calculation in which both 
the coefficients and the matrix elements of the operators are computed
with the same renormalization, a preferred scheme {\it does not} exist. 

\section{Anomalous dimensions at one and two loops}
\label{sec:andim}
In this section we recall the procedure for the calculation
of the anomalous dimension matrix $\hat \gamma$
in dimensional regularization. The section is divided in two parts:
in the first part,
we introduce the general formulae which define the anomalous dimension
matrix, in the second we give the practical recipe to compute it at the
NLO.

\subsection{General definitions and scheme dependence}
\label{subsec:gdsd} 
The ADM for the operators appearing in the effective theory is given by 
the matrix
\be
\hat\gamma =2 \, \Z^{-1}\mu^2\frac{d}{d\mu^2}\Z
\, , \label{gamma4}
\ee
where $\Z\equiv \Z(\as, \lambda)$, defined by the relation
\be \vec Q = \Z^{-1} \vec Q^B \, , \label{zd} \ee
 gives the renormalized operators in terms of the bare ones.
$\lambda$ is the renormalized gauge parameter on which,
in general, $\Z$ may depend, see for example ref.~\cite{Altarelli}. Note that in 
refs.~\cite{bw}--\cite{Ciuchini} the dependence on the gauge
parameter was ignored because, in $\MSbar$ schemes, $\Z$ is 
gauge independent. In this work, since we will compare 
anomalous-dimension matrices between schemes in which $\Z$ can be
gauge dependent, such as the RI scheme, this dependence has to be taken
explicitly into account.

In dimensional regularization, using eq.~(\ref{gamma4}),
one gets
\be
\hat\gamma= 2 \, \Z^{ -1} \left[ \left( -
\epsilon \as + \beta(\as) \right) \frac{\partial}{\partial \as} \Z
+ \lambda \beta_\lambda(\as) \frac{\partial}{\partial \lambda}\Z
\right] \, ,\label{RGgamma}
\ee
where $\epsilon= (4-D)/2$.
$\beta(\as)$ and $\beta_\lambda(\as)$ are the $\beta$ functions 
which govern the evolution of the effective coupling constant
and renormalized gauge parameter $\lambda$, respectively
\be \mu^2 \frac { d \as} {d \mu^2} = \beta (\as ) \, , 
\,\,\,\,\,\,\,\, \mu^2 \frac { d \lambda} {d \mu^2} =
\lambda \beta_\lambda (\as )\label{rcc} \, , \ee
with
\be \beta(\as) = - \beta_0 \frac {\as^2} {4\pi}
- \beta_1 \frac {\as^3} {(4\pi)^2} + O(\as^4)
\, , \,\,\,\,\,\,\, \beta_\lambda(\as) = - 
\frac {\as} {4\pi}\beta^0_\lambda +O(\as^2) \, . \ee
$\beta_0$, $\beta_1$ and $\beta^0_\lambda$ are given by
\bea \label{eq:betas} \beta_0 &=&\frac {(11N-2 n_f)} {3} \, , 
\,\,\,\,\,\,\, 
\beta_1 = \frac {34}{3} N^2 - \frac{10}{3} N n_f -\frac
{(N^2-1)}{N} n_f \, , \nn \\
\beta^0_\lambda &=& - \frac N2\left( \frac {13}{3}-\lambda \right)
+\frac{2}{3} n_f \, , \eea
where $n_f$ is the number of active flavours. The strong coupling 
constant $\as$ and the gauge parameter $\lambda$ are renormalized in
the $\MSbar$ scheme. This does not imply
that also the operators must be computed in $\MSbar$, because
the definition of the composite operators is an independent step 
that has nothing to do with the procedure
which renormalizes the parameters of the strong-interaction Lagrangian. 

From eq.~(\ref{RGgamma}), by writing $\hat\gamma$ and $\Z$ as series 
in the strong coupling constant
\be \hat\gamma =\frac{\as}{4\pi}\, \hat\gamma^{(0)}
 +\frac{\as^{ 2}}{(4\pi)^{ 2}} \, \hat\gamma^{(1)}
+ \cdots \, ,\label{gexp}\ee
and 
\be
\Z=1+\frac{\as}{4\pi}\Z^{ (1)}+
\frac{\as^{2}}{(4\pi)^2}\Z^{ (2)}+\cdots \, , 
\label{Zexpansion}
\ee
 we derive the following relations
\be
\hat\gamma^{ (0)}=-2\epsilon\Z^{ (1)}
\label{gamma0}
\ee
and
\be
\hat\gamma^{ (1)} =-4\epsilon\Z^{ (2)}-2 \beta_0\Z^{ (1)}+
2\epsilon\Z^{ (1)}\Z^{ (1)} -2 \beta_0\Z^{ (1)} -2 \beta^0_\lambda
\lambda \frac{\partial Z^{ (1)}}{\partial \lambda} 
\, .\label{gamma1}
\ee
We can expand $\Z^{(i)}$ in eqs.~(\ref{gamma0}) and (\ref{gamma1})
in inverse powers of $\epsilon$
\be
\Z^{ (i)}=\sum_{j=0}^{i}\left(\frac{1}{\epsilon}\right)^{ j}
\Z^{ (i)}_{ j}\ .
\label{Zpoleexp}
\ee

The requirement that anomalous dimension is finite as $\epsilon \rightarrow
0$ implies a relation between the one- and two-loop
coefficients of $\Z$ (note that in all the regularizations
$ \Z^{ (1)}_{ 1}$ is gauge invariant for gauge invariant operators)
\be
4\Z^{(2)}_{ 2}+2 \beta_0 \Z^{ (1)}_{ 1}
-2\Z^{ (1)}_{ 1}\Z^{ (1)}_{ 1}=0 \, ,
\label{twoloop_ad_condition}
\ee
which can be used as a check of the calculations. In addition, 
from the eqs.~(\ref{gamma0}) and (\ref{gamma1}) we obtain
\be \hat \gamma^{(0)}= -2 \Z^{(1)}_1 \label{1loopADM} \ee
and
\be
\hat\gamma^{ (1)}=-4\Z^{ (2)}_{ 1}-2 \beta_0 \Z_{
0}^{ (1)}+2(\Z^{ (1)}_{ 1}\Z^{ (1)}_{ 0}+
\Z^{ (1)}_{ 0}\Z^{ (1)}_{ 1})-2\beta
_\lambda ^0\lambda \frac{\partial Z_0^{(1)} }{\partial \lambda} \, .
\label{2loopADM}
\ee
Thus, it is sufficient to compute the pole and
finite part of $\Z^{(1)}$ and the single pole of $\Z^{(2)}$,
together with $\beta_0$ and $\beta^0_\lambda$, in order to obtain 
the two-loop anomalous dimension. Note that the last term in 
eq.~(\ref{2loopADM}) is absent in refs.~\cite{bw}--\cite{Ciuchini}.
Eq.~(\ref{2loopADM}) tells us how to derive $\hat \gamma^{(1)}$.
In dimensional regularizations, such as HV, NDR or DRED, the calculation is
complicated by the presence of the so-called ``effervescent" operators, 
which appear in the intermediate steps of the calculation \cite{Altarelli,bw}.
The EOs are independent operators which are present in D dimensions
but disappear in the physical basis of the 4-dimensional operators.
Because of the presence of the EOs, the products of the matrices
$\Z^{(i)}_j$ in eq.~(\ref{2loopADM}) have to be done by summing indices
over the full set of operators, including the EOs. Only
at the end of the calculation we can restrict the set of operators
to thoses of the physical 4-dimensional basis. As explained below, 
the identification of the EOs, and of the corresponding mixing matrix, 
can be completely avoided in RI schemes.

\subsection{Extraction of the the one- and two-loop anomalous dimension matrix}
\label{subsec:eotladm}
We now derive the general expression of the coefficients $Z_j^{(i)}$s in 
an arbitrary renormalization scheme, as obtained by using dimensional 
regularization. The derivation is general and, with trivial modifications, holds 
also with other regularizations, such as for example the lattice 
one~\cite{Ciuchini2}. Let us consider the matrix elements of generic bare 
operators, denoted 
as $\vec Q_B$, computed in a covariant gauge, between assigned quark and gluon 
external states. 
We define the matrix elements of $\vec Q_B$ as the 1PI
bare Green functions $\Gamma_{Q_B}$ multiplied by the renormalization constants
of the external fields 
\be \langle \vec Q_B\rangle = Z_\psi^{-2} \Gamma_{Q_B} \ ,  \ee
where $Z_\psi$ will be defined below.
 By calling $\alpha_0$ the dimensionless bare coupling
constant, for $p^2=-\mu^2$, where $p^2$ denotes 
generically the squared momentum of the external states,
we have~\footnote{ When working in the $\MSbar$ scheme, 
it is convenient to express
the poles  in terms of $1/\bar \epsilon=1/\epsilon -\gamma_E+\ln(4 \pi)$,
where $\gamma_E$ is the Euler gamma. The formulae below are valid also in the
$\MSbar$ scheme if one interprets $1/\epsilon$ as $1/\bar \epsilon$.} 
\bea
\label{Obare}
\langle \vec Q_B\rangle &=&
\left[ 1+\frac{\alpha _0}{4\pi }
\left(\hat A_0+\frac{\hat A_1}{\epsilon} \right) 
 \right. \nn \\ 
&+& \left. 
\left( \frac{\alpha _0}{4\pi }\right)^2
\left(\hat B_0+\frac{\hat B_1}\epsilon +\frac{\hat B_2}{\epsilon ^2}\right) 
\right] \langle \vec Q^{(0)}\rangle \, . 
\eea
 $\langle \vec Q^{(0)} \rangle$ are the tree-level matrix elements of
all the operators of the regularized theory, including the EOs.
At one loop, for gauge-invariant operators, the matrix $\hat A_1$ is gauge 
and regularization independent, whilst $\hat A_0$ can be written in the form
\begin{equation}
\label{A0-lam}
\hat A_0(\lambda _0)=\hat A_0(0)+\lambda _0\frac{\partial \hat A_0}{\partial
\lambda _0} \, 
\end{equation}
where $\lambda_0$ is the bare gauge-parameter. Note that also
$\partial \hat A_0 /\partial \lambda _0$ is regularization independent.

In eq.~(\ref{Obare}),
we substitute the bare parameters $\alpha _0$ and
$\lambda _0$ with their renormalized counter-parts, according to
\begin{equation}
\label{lambda0}\lambda _0=\lambda \left( 1-\frac{\alpha _s}{4\pi }
\frac{\beta _\lambda ^0}\epsilon +\ldots \right) \, , 
\,\,\,\,\,\,\,\, 
\alpha _0=\alpha _s\left( 1-\frac{\alpha _s}{%
4\pi }\frac{\beta _0}\epsilon +\ldots \right) \, . 
\end{equation}
At the NLO, and taking into account that $\hat A_1$ is gauge invariant,
we can ignore all other terms which relate the bare and the
renormalized coupling constant and gauge parameter.

For a given, generic renormalization scheme, we can write the following
relation between matrix elements
\begin{equation}
\label{Orin} \langle Q_R\rangle =\Z^{-1} \langle Q_B\rangle =
\left( 1+\frac{\alpha_s}{4\pi } \hat r \right) \langle Q^{(0)}\rangle \, ,
\end{equation}
where the matrix $\hat r$ defines the renormalization scheme.
With a little algebra, a comparison 
of eqs.~(\ref{Obare}), (\ref{lambda0}) and (\ref{Orin}) gives the mixing matrix $\Z$ in terms
of $\hat A_0$, $\dots$ $\hat B_2$ (we list only the terms which are needed
at the NLO)
\bea\label{zhh} 
\hat Z_0^{(1)}&=&\hat { A}_0-\hat r\quad ,\quad \hat Z_1^{(1)}=\hat 
{ A}_1 \, , \nn \\ \hat Z_1^{(2)}&=&\hat { B}_1-\hat { A}_1 \hat r-\beta
_0 \hat { A}_0-\beta _\lambda ^0\lambda 
\frac{\partial \hat { A}_0}{\partial \lambda } \, , \\ 
\hat Z_2^{(2)}&=&\hat { B}_2-\beta _0 \hat { A}_1 \, . \nn 
 \eea
$\hat \gamma ^{(1)}$ is then readily obtained
by substituting the relations (\ref{zhh}) in eq.~(\ref{2loopADM}). 
To this purpose, we express the regularization and 
renormalization-independent combination (obviously $\hat \gamma^
{(0)} =-2 \hat { A}_1$)
\be \hat G = \hat \gamma^{(1)} - \Bigl[ \hat r , \hat \gamma^
{(0)} \Bigr] - 2 \beta_0 \hat r -2\beta
_\lambda ^0\lambda \frac{\partial \hat r}{\partial \lambda } \label{hrb} \ee
in terms of the matrices $\hat A_0$, $\dots$ $\hat B_2$,
\begin{equation}
\hat G= -4 \left[ \hat B_1 - \frac{1}{2} \left( \hat A_1
\hat A_0+\hat A_0 \hat A_1\right) -\frac{1}{2} \beta_0
\hat A_0-\frac{1}{2} \beta_\lambda^0
\lambda \frac{\partial \hat A_0}{\partial \lambda }\right] \, .
\label{eq:ahah}
\end{equation}
Equation~(\ref{eq:ahah}) demonstrates that $\hat G$ is renormalization-scheme 
independent since the r.h.s. does not depend on $\hat r$. $\hat G$ is also 
regularization independent as the following, simple argument demonstrates.
Let us compute the matrix elements of the renormalized operators using two 
different regularizations, but the same external quark and gluon states and 
in the same gauge. Irrespectively of the regularization used in the calculations, 
we can define the operators in the same renormalization scheme in terms of 
the mixing matrix $\hat r$ in eq.~(\ref{Orin}). Since the renormalized 
operators are the same, they obey the same renormalization-group equations. 
Thus, not only $\hat r$, but also $\hat \gamma^{(1)}$ is the same in the two 
cases. This demonstrates that the l.h.s. of eq.~(\ref{eq:ahah}) is also
regularization independent.

Note that the last term of eq.~(\ref{hrb}) is absent in 
refs.~\cite{bw}--\cite{Ciuchini}. This is because in all $\MSbar$ schemes
the derivative $\partial \hat r / \partial \lambda$ is regularization 
invariant, i.e. it is the same for two different $\MSbar$ regularizations. 
Thus, in these schemes, the difference between the two-loop ADMs 
is given by:
\be
\label{delta_g1}
\Delta \hat \gamma^{(1)} = \Bigl[ \Delta \hat r , \hat \gamma^
{(0)} \Bigr] + 2 \beta_0 \Delta \hat r
\ee
On the other hand, the combination $\hat G$ depends on the external states 
used in the calculation, and on the gauge, because $\hat r$ depends on 
these variables. The RI scheme is defined, for given external states and
at a fixed gauge, by the condition $\hat r=0$. Thus, in this scheme, $\hat G$ 
coincides with the two-loop anomalous dimension. 

Equation~(\ref{eq:ahah}) provides also a practical method to compute the 
ADM in terms of the one-loop matrices 
$\hat A_1$ and $\hat A_0$ and of the two-loop pole term $\hat B_1$. 
In refs.~\cite{bw} and \cite{Ciuchini}, it was demonstrated that the 
equation which allows to compute $\hat\gamma^{(1)}$ in terms of the one- 
and two-loop renormalization matrices is valid diagram-by-diagram.
In ref.~\cite{Ciuchini} it was also shown that the relations between 
the anomalous dimensions in different regularizations/renormalizations 
can also be established on a diagram-by-diagram basis. Using these 
observations, and eq.~(\ref{eq:ahah}), we give the recipe to obtain in 
a very simple way the anomalous dimension matrix in the RI scheme, 
$\hat \gamma^{(1)}_{RI}$:
\begin{itemize} 
\item[i)] 
Choose the set of external states, and the gauge, which define the RI 
scheme of interest.
\item[ii)] 
Compute a given two-loop diagram where the bare operator is inserted.
\item[iii)] 
Subtract to it the result obtained by substituting, to any 
internal subdiagram, one half of the amplitude of the corresponding one-loop 
diagram computed at $p^2=-\mu^2$:
\be
\frac{1}{2} \frac{\as}{4\pi} 
\left(\hat A_0 +\frac{\hat A_1}{\epsilon} \right)
\langle \vec Q^{(0)}\rangle \, , 
\ee
i.e. one half of the contribution of the subdiagram to $\hat { A}_0$
and $\hat { A}_1$.
\item[iv)] 
When the internal subdiagram contributes to the renormalization of $\as$ 
and of the gauge parameter $\lambda$, apply iii) by inserting, in the two-loop 
diagram, only the divergent part of one-loop diagram (always with a factor 1/2). 
This rule corresponds to the choice of renormalized $\as$ and $\lambda$ in 
the $\MSbar$ scheme.
\item[v)] The coefficient of the single pole obtained from steps i) - iv) 
is the contribution of the given two-loop diagram to the combination 
(\ref{eq:ahah}).
\end{itemize}
With this procedure, we do not need to isolate the EOs from the operators
of the 4-dimensional basis. The reason is two-fold. On the one hand, in the 
RI scheme, the counterterms corresponding to the EOs and to the operators 
of the 4-dimensional basis (i.e. the combination $\hat A_1 \hat A_0+\hat A_0 
\hat A_1$ of eq.~(\ref{eq:ahah})) are both subtracted with the same factor 1/2.
As shown in eq.~(\ref{g1-ms}) below, in the $\MSbar$ scheme, instead, different 
factors, namely 1 and 1/2, enter  the subtraction for the  4-dimensional and 
effervescent counterterms. Moreover, the subtracted diagrams, as obtained from 
steps i) to v), only contain simple poles or finite terms, while the double 
poles completely cancel out. Thus, the projection on the physical 4-dimensional 
basis cannot give rise to further single-pole terms due to the EOs.

For completeness, we now give the definition of the quark wave-function 
renormalization which we used in the RI scheme. We introduce the two-point Green 
function, computed in the same gauge as the RI scheme at hand,
\be \Gamma_{\psi}(p^2) = \frac{i}{48}{\rm tr} 
\Bigl( \gamma ^{\mu} \frac{\partial S(p)^{-1}}{\partial p^{\mu} } 
\Bigr) \, , \label{eq:zpsiq} \ee
where the trace is taken over colour and spin indices.
The renormalization condition for the quark fields is given by
\be 
\label{eq:zpsir}
Z_\psi^{-1} (\mu^2) \Gamma_{\psi}(p^2=\mu^2)=1 \, . 
\ee
Note that, in the calculation of the four-point Green functions, in any given 
scheme, different choices of the wave-function renormalization correspond to 
different choices of the quark external states. Thus, they also imply,
in practice, different definitions of the renormalized operators. Obviously,
all these choices are equivalent in principle, and they do not affect the 
calculation of the physical quantities at the order  we are working.
In the RI scheme, however, the specific choice of eq.~(\ref{eq:zpsir}) has 
the advantage that the vector and axial-vector currents, renormalized according
to the same rules used for the four-fermion operators, satisfy automatically 
the relevant Ward identities. This is true for all the regularizations used 
in the intermediate steps and thus provides a useful check of the calculations. 
The validity of the Ward identities among renormalized quantities is not 
a-priori guaranteed, and it does not occur, for instance, in the 
HV- or DRED-$\MSbar$ schemes, because of the chiral symmetry breaking induced 
by the regularization. In the latter cases, the finite one-loop coefficient, 
entering  the forward matrix element of the axial-vector current, does 
not vanish. These finite corrections are compensated, in the evolution 
equation of the Wilson coefficients, by a term appearing in the two-loop 
current anomalous dimension.

In order to obtain the anomalous dimension in the $\MSbar$ scheme, one can 
proceed in two ways. The first was explained in refs.~\cite{bw}--\cite{Ciuchini} 
and the details will not be given here. It is based on the relation
\begin{equation}
\label{g1-ms}
\hat \gamma _{\overline{{\rm MS}}}^{(1)}=-4\left[\hat B_1-
\left(\bar A_1\right)\left(\bar A_0\right)-\frac {1}{2}\left(\widetilde 
A_1\right)\left(\widetilde A_0\right)-\beta _0\bar A_0-\beta _\lambda
^0\lambda \frac{\partial \bar A_0}{\partial \lambda }\right] \, .
\end{equation}
which can be derived from eqs.~(\ref{hrb}) and (\ref{eq:ahah}) by using 
$\hat r_{\msbar}= \bar A_0$. In eq.~(\ref{g1-ms}), we denoted as $\bar A_i$ 
the matrix elements restricted to the operators of the four-dimensional basis, 
and as $\widetilde A_i$ those connecting the operators of the four-dimensional 
basis with the effervescent ones. $\hat \gamma _{\overline{{\rm MS}}}^{(1)}$ 
is obviously the restricted matrix. The second method to obtain $\hat 
\gamma _{\overline{{\rm MS}}}^{(1)}$, is by using the relation
\begin{equation}
\label{check1}
\hat \gamma _{RI}^{(1)}=\hat \gamma _{\overline{{\rm MS}}}^{(1)}-2\left(
{\bar A}_1 {\bar A}_0-{ \bar A}_0 { \bar A}_1\right) -2\beta _0
{ \bar A}_0-2\beta _\lambda ^0\lambda 
\frac{\partial { \bar A}_0}{\partial \lambda }
\end{equation}
which follows from eqs.~(\ref{eq:ahah}) and (\ref{g1-ms}) and it is also 
valid diagram by diagram. Note that the change of scheme in eq.~(\ref{check1})
is equivalent to the substitution
\be \J_{\msbar} = \J_{RI} - \hat r_{\msbar}^T \ee
discussed in eq.~(\ref{eq:jri}) of subsec.~\ref{subsec:cf}.

\subsection{Checks of the calculation}
\label{subsec:check}
In this subsection, we illustrate several checks that have been made in
order to verify the correctness of our calculations:
\begin{itemize}
\item in the NDR-$\MSbar$ renormalization scheme, the ADM of the 
operators $Q_1^{\pm}$, $Q_2^\pm$ and $Q_3^\pm$ can be extracted from
the results of refs.~\cite{bjl,Ciuchini}. They agree with the results 
presented in this paper;
\item we have computed the anomalous dimension matrix both in $\MSbar$
and in RI and verified that the result satisfy eq.~(\ref{hrb}), i.e. that we get exactly the same 
$\hat G$ in the two schemes;
\item by computing the two-loop diagrams both in the $\MSbar$ and in the
FRI schemes, we verified that the relation in eq.~(\ref{check1}) holds, as 
we mentioned above, diagram by diagram~\cite{Ciuchini}. 
\end{itemize}

\section{The anomalous dimension matrix}
\label{sec:admf}
In this section, we give the results for the anomalous dimension matrix in the
Feynman-gauge RI scheme, which will be defined precisely in 
subsec.~\ref{subsec:sec}, 
and the matrices necessary to pass  from FRI to 
a) NDR-$\MSbar$, as defined in refs.~\cite{bw}--\cite{Ciuchini};
b)  the Landau-gauge RI scheme, which is the most suitable 
for the calculation of the matrix elements on the lattice, using operators 
renormalized non-perturbatively~\cite{NP}--\cite{JAPBK}. The FRI scheme is 
presented only because it is the simplest for doing the perturbative 
calculations. In practical cases, we expect that the $\MSbar$ and LRI schemes 
will be used for phenomenological applications.

\subsection{The anomalous dimension matrix at LO and at the 
NLO in FRI}
\label{subsec:sec}
In this subsection, we present the results of the leading order ADM
and of the next-to-leading order ADM in the FRI scheme. 
\par
\begin{figure}
\begin{center}
\epsfxsize=\textwidth
\leavevmode\epsffile{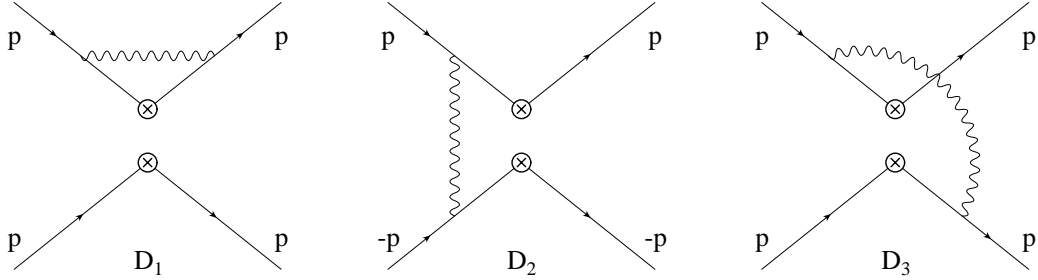}
\caption[]{\it One-loop Feynman diagrams. We show the external quark momenta
chosen to obtain the results in the
FRI scheme. In the LRI scheme,
all the external momenta are equal to $p$.}
\protect\label{fig:ol}
\end{center}
\end{figure}
\begin{figure}
\begin{center}
\epsfxsize=\textwidth
\leavevmode\epsffile{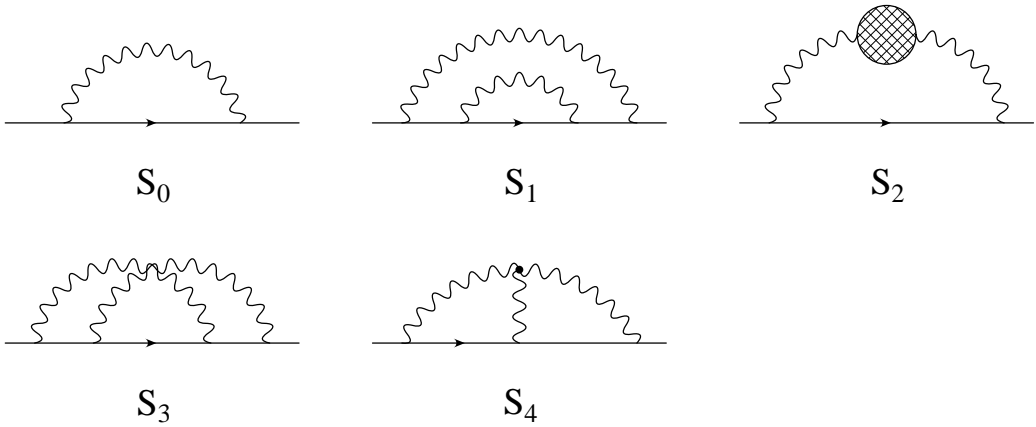}
\caption[]{\it One- and two-loop diagrams for the quark propagator.}
\label{fig:se}
\end{center}
\end{figure}

The results in FRI have been obtained by computing the one- and two-loop
Feynman diagrams shown in figs.~\ref{fig:ol}-\ref{fig:se} in the Feynman 
gauge. At one loop, we have taken the external quark momenta as indicated in 
fig.~\ref{fig:ol}; in the two-loop case, when the external subdiagrams are 
$D_1$, $D_2$ or $D_3$, the external momenta have been chosen as the 
corresponding ones in fig.~\ref{fig:ol}.
The field renormalization constant is  computed
in  FRI according to eq.~(\ref{eq:zpsir}).
\par  Although the results in RI only 
depend on the external momenta and the gauge, but not on the regularization,
we specify that we did the calculation using NDR. The choice 
of the external momenta may appear rather strange, since it is different for
the different one-loop diagrams. It is, however, particularly convenient for 
the perturbative calculation. In the $\MSbar$ scheme, the results for the ADM 
are not affected, since the independence  of the external states
is valid diagram-by-diagram.
In tables~\ref{tab:poli} and \ref{tab:polis} we give the complete list of 
the single poles necessary to compute the one- and two-loop anomalous 
dimension matrix. 
\begin{figure}
\begin{center}
\epsfxsize=\textwidth
\leavevmode\epsffile{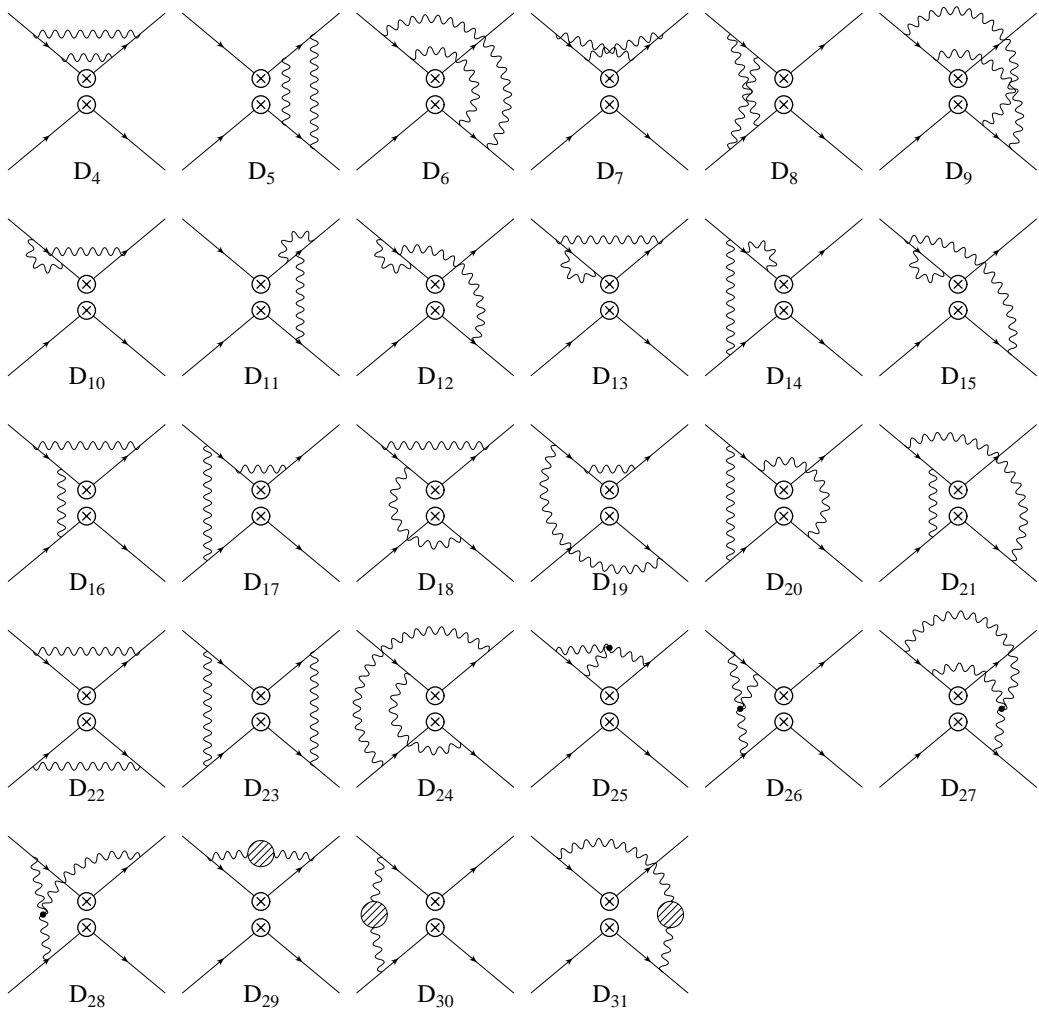}
\caption[]{\it Two-loop Feynman diagrams.}
\label{fig:tl}
\end{center}
\end{figure}

{\scriptsize\begin{table}
\begin{center}
\begin{tabular}{|c|c|c|c|c|c|c|c|c|}\hl
Diag. & Mult. & $1\ra 1$ & $2\ra 2$ & $3\ra 3$ & $4\ra 4$ &
 $4\ra 5$ & $5\ra 4$ & $5\ra 5$ \\ \hl
1  &2   &  1  &  1 &  4 & 4  &  0  &  0  & 0  \\ 
2  &2   & -4  &  -1  & -1  & -1  & -1/4  & -12  & -3  \\ 
3  &2   &  1  &  4  &  1  &  1  &  -1/4  &  -12  & 3 \\ 
4  &2   & 5/4  & 5/4  &  8  & 8  & 0  & 0  & 0  \\ 
5  &2   &  8  & 5/4  & 5/4  &  2  & 1/2  & 24  &  6  \\
6  &2   & 5/4  &  8  & 5/4  &  2  & -1/2  & -24  &  6  \\
7  &2   & -2  & -2  & -2  &  -2  & 0  & 0  &  2  \\
8  &2   & -2  & -2  & -2  & 1  & -1/4  & -12  & -1  \\
9  &2   &  -2  & -2  & -2  & 1  & 1/4  & 12  & -1  \\
10  &4   &  1  &  1  & 1  &  1  &  0  &  0  & 1  \\
11  &4   &  -1  & -1  & -1  &  -1  & 0  & 0  & -1  \\
12  &4   &  1  & 1  &  1  &  1  & 0  & 0  & 1  \\
13  &4   &  -1  & -1  & -4  &  -4  & 0  & 0  & 0  \\
14  &4   &  4  & 1  &  1  &  1  & 1/4  & 12  &  3  \\
15  &4   & -1  & -4   & -1  & -1  & 1/4  &  12  & -3  \\
16  &4   & 0  & -3/4  & 0  &  0  &  0  & 0  &  0  \\
17  &4   &  0  & -3/4  & 0  &  0  &  0  &  0  &  0  \\
18  &4   & 3/4  & 0  &  0  &  0  &  0  & 0  & 0  \\
19  &4   &  3/4  & 0  &  0  &  0 &  0  &  0  &  0  \\
20  &4   &  0  & 0  & -3/4  & 0  & 0  & 0  & 0  \\
21  &4   &  0  &  0  & -3/4  &  0  & 0  & 0  & 0  \\
22  &1   &  0  &  0  & 0  &  0  &  0  & 0  &  0  \\
23  &1   &  0  &  0  & 0  &  0  &  0  & 0  &  0  \\
24  &1   &  0  &  0  &  0  &  0 &  0  &  0 &  0  \\
25  &4   &  -7/2  & -7/2  & -14  & -14  & 0  & 0  & 0  \\
26  &4   &  14  &  7/2  & 7/2  & 7/2  & 7/8  & 42  & 21/2  \\
27  &4   &  -7/2  & -14  & -7/2  & -7/2  & 7/8  & 42  & -21/2  \\
28  &4   &  0  &  0  & 0  &  0  &  3/4  &  -36  & 0  \\
29 $(N_c)$&2   &  5/4  &  5/4  &  23/3   & 23/3  & 0  & 0  & -8/9  \\
29 $(n_f)$&2   &  -1/2  &  -1/2  & -8/3  &  -8/3  & 0  & 0  &  2/9  \\
30 $(N_c)$&2   &  -23/3  & -5/4  &  -5/4 &  -5/4  & -77/144  & -77/3 & -199/36  \\
30 $(n_f)$&2   &  8/3  &  1/2  & 1/2    &  1/2  & 13/72  &  26/3  & 35/18  \\
31 $(N_c)$&2   &  5/4  &  23/3  & 5/4   &  5/4  &  -77/144  &  -77/3  & 199/36 \\
31 $(n_f)$&2 
  &  -1/2  &  -8/3  & -1/2  &  -1/2  & 13/72  &  26/3  & -35/18  \\
\hl
\end{tabular}
\caption{\it Single pole contributions of the one- and
two-loop diagrams to the ADM
in the FRI scheme. In this scheme the double poles are absent.
The label $i \ra j$ denotes the indices of the mixing-matrix.
Thus $4 \ra 5$ corresponds to $[\hat Z^{(2)}_1]_{45}$,
i.e. the mixing of the bare operator $Q_4$ with $Q_5$, without colour factors.
 The first column
refers to the diagram labels defined in fig.~\protect\ref{fig:ol} and
\protect\ref{fig:tl}, the second 
column to the diagram multiplicity. For the last three diagrams, 
we indicate separately the term proportional to $N_c$ or $n_f$ coming from
the gluon vacuum-polarization of the internal gluon line,
 see fig.~\protect\ref{fig:tl}.} 
\label{tab:poli}
\end{center}
\end{table}}
\begin{table}
\begin{center}
\begin{tabular}{|c|c|c|c|c|c|}\hl \hl
$S_0$ & $S_1$ & $S_2$ $(N_c)$ & $S_2$ $(n_f)$ & $S_3$ & $S_4$ \\ \hl
 -1 & 3/4 & 5/4 & 1/2 & 0 & 7 \\ 
\hl \hl
\end{tabular}
\caption{\it Single pole contributions to the ADM of the one- and 
two-loop self-energy diagrams in the FRI scheme. For $S_2$ the 
contributions proportional to $N_c$ (3rd column) and $n_f$ (4th column)
are shown separately.}
\label{tab:polis}
\end{center}
\end{table}

In order to present results for the ADM, we expand the coefficients of 
$\hat \gamma^{\pm}_{FRI}$, written as  in eq.~(\ref{str1}), in powers of $\as$
\bea 
& & A^{\pm} =  \frac{\as}{4\pi} A^{\pm}_1 + \frac{\as^2}{(4\pi)^2} A^{\pm}_2 
+ \dots \nn \\ 
& & B =  \frac{\as}{4\pi} B_1 + \frac{\as^2}{(4\pi)^2} B_2 + \dots \\ 
& & \vdots \nn \\
& & I^{\pm} =  \frac{\as}{4\pi} I^{\pm}_1 + \frac{\as^2}{(4\pi)^2} I^{\pm}_2 
+ \dots \nn 
\eea
and give the expression for these quantities.
 
 \par
At one-loop the ADM is independent of renormalization scheme, 
external states and gauge. The results in this case are the following
\renewcommand{\arraystretch}{1.5}
\be
\brr{l c l}
A^+_1=6-\dfrac{6}{N_c} \:\: &\ & \:\: A^-_1=-6-\dfrac{6}{N_c} \cr
B_1=\dfrac{6}{N_c} \:\: &\ & \:\: C_1=12 \cr
D_1=0 \:\: &\ & \:\: E_1=-6N_c+\dfrac{6}{N_c} \cr
F^+_1=6-6N_c+\dfrac{6}{N_c}\:\: &\ & \:\:F^-_1=-6-6N_c+\dfrac{6}{N_c} \cr
G^+_1=\dfrac{1}{2}-\dfrac{1}{N_c}\:\: &\ & \:\:G^-_1=-\dfrac{1}{2}-
 \dfrac{1}{N_c} \cr
H^+_1=-24-\dfrac{48}{N_c}\:\: &\ & \:\:H^-_1=24-\dfrac{48}{N_c} \cr
I^+_1=6+2N_c-\dfrac{2}{N_c}\:\: &\ & \:\:I^-_1=-6+2N_c-\dfrac{2}{N_c}\ .
\err
\ee
\renewcommand{\arraystretch}{1.0}
Our results agree with those obtained in ref.~\cite{beautysusy2}.
\par
For the two-loop ADM we obtained (with $\beta_\lambda^0$  computed from
eq.~(\ref{eq:betas}) for $\lambda=1$)  
\bea
A_2^{\pm} &=& 
-{{209}\over 3} - {{57}\over {2\,{N_c ^2}}} \pm {{39}\over {N_c }} 
\pm {{355\,N_c }\over 6} \mp {{32\,n_f}\over 3} + 
{{32\,n_f}\over {3\,N_c }} \pm 3\, \beta_\lambda^0 \left( 1 \mp 
{1\over {N_c }}\right) \nn \\
B_2  &=& 
{{355}\over 6} + {{15}\over {2\,{N_c ^2}}} - {{32\,n_f}\over {3\,N_c }} 
+ {{3\, \beta_\lambda^0} \over {N_c }} \nn \\
C_2  &=& 
- {6\over {N_c }} + {{418\,N_c }\over 3} - {{64\,n_f}\over 3} 
+ 6\,\beta_\lambda^0 \nn \\
D_2  &=& 
{9\over {N_c }} - {{9\,N_c }\over 4} \nn \\
E_2  &=& 
{{481}\over 6} + {{15}\over {2\,{N_c ^2}}} - {{445\,{N_c ^2}}\over 6} 
- {{32\,n_f}\over {3\,N_c }} + {{32\,N_c \,n_f}\over 3}
+ 3\, \beta_\lambda^0 \left({1\over {N_c }} - N_c \right) \\
F_2^{\pm} &=& 
{{209}\over 3} - {{27}\over {2\,{N_c ^2}}} \pm {{272\,N_c }\over 3} 
- {{445\,{N_c ^2}}\over 6} \mp {{32\,n_f}\over 3} - 
{{32\,n_f}\over {3\,N_c }} + {{32\,N_c \,n_f}\over 3} \nn \\
& & \pm 3\, \beta_\lambda^0 \left( 1 \pm {1\over {N_c }} \mp N_c \right)
\nn \\
G_2^{\pm} &=& 
-{{263}\over {18}} - {2\over {{N_c ^2}}} \pm {4\over {N_c }} \pm 
{{59\,N_c }\over 9} \mp {{8\,n_f}\over 9} + {{16\,n_f}\over {9\,N_c }} 
\pm {1\over 4} \, \beta_\lambda^0 \left( 1 \mp {2\over {N_c }}\right) \nn \\
H_2^{\pm} &=& 
-{{1240}\over 3} - {{96}\over {{N_c ^2}}} \pm {{192}\over {N_c }} \mp 
{{800\,N_c }\over 3} \pm {{128\,n_f}\over 3} + {{256\,n_f}\over {3\,N_c }}
\mp 12\, \beta_\lambda^0 \left(1 \pm {2\,\over {N_c }}\right) \nn \\
I_2^{\pm} &=& 
-{{209}\over 9} - {{59}\over {2\,{N_c ^2}}} \pm {{32}\over {N_c }} \pm 
{{140\,N_c }\over 3} + {{409\,{N_c ^2}}\over {18}} \mp {{32\,n_f}\over 3} + 
{{32\,n_f}\over {9\,N_c }} - {{32\,N_c \,n_f}\over 9} \nn \\
& & \pm \beta_\lambda^0 \left( 3 \mp {1\over {N_c }} \pm N_c \right) \nn
\ .\eea 
For $\gamma^{\pm \, (1)}_{FRI}$, we have shown explicitely those terms, 
proportional to $\beta_\lambda^0$,  which cancel 
$\lambda \partial \J_{FRI} / \partial \lambda$ in eq.~(\ref{jj1}).
From the one- and two-loop matrix elements of $\hat \gamma^{\pm}_{FRI}$, 
by solving eq.~(\ref{jj1}), one can easily compute $\J_{FRI}$.
By writing $\J_{FRI}$ as  
\be 
\J^{\pm}_{FRI} \equiv \pmatrix{
J_{11}^{\pm} & 0 & 0 & 0 & 0\cr
0 & J_{22} & \pm J_{23} & 0 & 0\cr
0 & \pm J_{32} & J_{33} & 0 & 0\cr
0 & 0 & 0 & J_{44}^{\pm} & J_{45}^{\pm} \cr
0 & 0 & 0 & J_{54}^{\pm} & J_{55}^{\pm} \cr } \, , 
\label{str2}
\ee
we obtain
\bea
\nn J^+_{11}&=&\frac{ 
   -\left( 23931 - 2862\, n_f + 128\, n_f^2 \right) }{ 
    6\left( 33 - 2\, n_f \right) ^2 } \\
\nn J^-_{11}&=&\frac{ 
   28089 - 3114\, n_f + 128\, n_f^2}{ 
    3\left( 33 - 2\, n_f \right) ^2 }\\
\nn J_{22}&=&\frac{ 
   -\left( 1437345 - 221058\, n_f + 13488\, n_f^2 - 
      256\, n_f^3 \right) }{ 
    24\left( 33 - 2\, n_f \right) ^2
     \left( 30 - n_f \right) }\\
\nn J_{23}&=&\frac{ 
   45}{ 16\left( 30 - n_f \right) }\\
\nn J_{32}&=&\frac{ 
   -4347675 + 2468583\, n_f - 294786\, n_f^2 + 
     14928\, n_f^3 - 256\, n_f^4}{ 
    4\left( 30 - n_f \right) \left( 3 - n_f \right) 
     \left( 33 - 2\, n_f \right) ^2 }\\
\nn J_{33}&=&\frac{ 
   -\left( -15575085 + 2142036\, n_f - 115572\, n_f^2 + 
      2048\, n_f^3 \right) }{ 
    24\left( 33 - 2\, n_f \right) ^2
     \left( 30 - n_f \right) }\\
\nn J^+_{44}&=&\frac{ 
   4176675 - 5048688\, n_f + 669548\, n_f^2 - 
     36624\, n_f^3 + 640\, n_f^4}{ 
    3\left( 33 - 2\, n_f \right) ^2
     \left( 125 - 132\, n_f + 4\, n_f^2 \right) }\\
  J^-_{44}&=&\frac{ 
   12084435 - 14286828\, n_f + 1744892\, n_f^2 - 
     86016\, n_f^3 + 1408\, n_f^4}{ 
    3\left( 33 - 2\, n_f \right) ^2
     \left( 125 - 132\, n_f + 4\, n_f^2 \right)} \\
\nn J^+_{45}&=&\frac{ 
   20\left( -277425 - 767424\, n_f + 118876\, n_f^2 - 
      7056\, n_f^3 + 128\, n_f^4 \right) }{ 
    3\left( 33 - 2\, n_f \right) ^2
     \left( 125 - 132\, n_f + 4\, n_f^2 \right) }\\
\nn J^-_{45}&=&\frac{ 
   4\left( 791235 + 1102188\, n_f - 152932\, n_f^2 + 
      7776\, n_f^3 - 128\, n_f^4 \right) }{ 
    3\left( 33 - 2\, n_f \right) ^2
     \left( 125 - 132\, n_f + 4\, n_f^2 \right) }\\
\nn J^+_{54}&=&\frac{ 
   -898695 + 1066800\, n_f - 142204\, n_f^2 + 
     7632\, n_f^3 - 128\, n_f^4}{ 
    36\left( 33 - 2\, n_f \right) ^2
     \left( 125 - 132\, n_f + 4\, n_f^2 \right) }\\ 
\nn J^-_{54}&=&\frac{ 
   5\left( 1208169 - 1422948\, n_f + 169780\, n_f^2 - 
      8064\, n_f^3 + 128\, n_f^4 \right) }{ 
    36\left( 33 - 2\, n_f \right) ^2
     \left( 125 - 132\, n_f + 4\, n_f^2 \right) }\\ 
\nn J^+_{55}&=&\frac{ 
   -11915775 + 14548416\, n_f - 2050844\, n_f^2 + 
     119952\, n_f^3 - 2176\, n_f^4}{ 
    9 \left( 33 - 2\, n_f \right) ^2
     \left( 125 - 132\, n_f + 4\, n_f^2 \right) }\\
\nn J^-_{55}&=&\frac{ 
   -957555 + 1949172\, n_f - 126428\, n_f^2 - 
     2016\, n_f^3 + 128\, n_f^4}{ 
    9 \left( 33 - 2\, n_f \right) ^2
    \left( 125 - 132\, n_f + 4\, n_f^2 \right) }\ .
\eea

\subsection{Relation between FRI and other renormalization schemes}
In this subsection, we give the recipe to pass from FRI to other schemes 
which may be useful for practical applications: LRI, for lattice calculations, 
and the standard $\MSbar$ NDR scheme. All we need to know is the shift matrix 
\bea 
\J_{LRI} &=& \J_{FRI} + \hat r^T_{FRI}
\nn \\ \J_{\msbar} &=& \J_{FRI} + \hat r^T_{FRI}-\hat r^T_{\msbar} \, .
\eea
From the knowledge of $\J$ in a given renormalization scheme, we can 
immediately obtain the evolution matrix $\W[\mu,M]$ using 
eqs.~(\ref{monster}) and (\ref{mo2}).

As discussed in subsec.~\ref{subsec:cf}, the renormalization scheme is 
completely defined by the matrix $\hat r$ of eq.~(\ref{Orin}), computed 
for given external momenta and gauge. 
We choose  quarks with equal momentum $p$ as external states and the Landau
gauge. The field renormalization constant is  computed
in  LRI according to eq.~(\ref{eq:zpsir}), which gives $Z_\psi=1$.
  We denote as $\hat r_{FRI}$, $\hat r_{LRI}$ and $\hat r_{\msbar}$
the three cases considered here: $\hat r_{LRI}$ is obviously zero and
$\hat r_{\msbar}$ is a $20 \times 20$ matrix because this regularization
(and consequently the corresponding $\MSbar$-renormalization scheme)
does not respect chiral and Fierz symmetries. By denoting with $\hat
r^{++}$ the $5 \times 5$ sub-matrix in the $Q^+_i$ sector and similarly
for $\hat r^{+-}$, $\hat r^{-+}$ and $\hat r^{--}$, we get 
(see also ref.~\cite{Ciuchini2} for the operators $Q^\pm_{1,2,3}$),
\bea
(\hat r^{\pm \pm}_{FRI})_{11} & = &
\pm \left( {3\over 2} + 12\,\ln 2 \right) - {3\over {2\,N_c}} - 
 {{12\,\ln 2}\over {N_c}} \nn \\
(\hat r^{\pm \pm}_{FRI})_{22} & = &
{1\over {2\,N_c}} - {{2\,\ln 2}\over {N_c}} \nn \\
(\hat r^{\pm \pm}_{FRI})_{23} & = & 
\pm \left( 1 - 4\,\ln 2 \right) \nn \\
(\hat r^{\pm \pm}_{FRI})_{32} & = & 
\mp \left( {1\over 2} + \ln 2 \right) \nn \\
(\hat r^{\pm \pm}_{FRI})_{33} & = &
{1\over {2\,N_c}} - {{2\,\ln 2}\over {N_c}} - 
 {{3\,N_c}\over 2} \\
(\hat r^{\pm \pm}_{FRI})_{44} & = &
\pm \left( {1\over 2} + 4\,\ln 2 \right) + {1\over {2\,N_c}} - 
 {{2\,\ln 2}\over {N_c}} - {{3\,N_c}\over 2} \nn \\
(\hat r^{\pm \pm}_{FRI})_{45} & = &
\pm \left( {5\over {24}} + {{2\,\ln 2}\over 3} \right) - 
{1\over {6\,N_c}} - {{5\,\ln 2}\over {6\,N_c}} \nn \\
(\hat r^{\pm \pm}_{FRI})_{54} & = &
\mp \left( 2 - 32\,\ln 2 \right) - {8\over {N_c}} - 
{{40\,\ln 2}\over {N_c}} \nn \\
(\hat r^{\pm \pm}_{FRI})_{55} & = &
\pm \left( {7\over 6} + {{28\,\ln 2}\over 3} \right) - {5\over {6\,N_c}} - 
 {{26\,\ln 2}\over {3\,N_c}} + {{N_c}\over 2} \nn
\eea
and
\bea
(\hat r^{\pm \pm}_{\msbar})_{11} & = &
\mp \left( 7 - 12\,\ln 2 \right) + {7\over {N_c}} - 
{{12\,\ln 2}\over {N_c}} \nn \\
(\hat r^{\pm \pm}_{\msbar})_{22} & = &
- {{2}\over {N_c}} - {{2\,\ln 2}\over {N_c}} \nn \\
(\hat r^{\pm \pm}_{\msbar})_{23} & = & 
\mp \left( 4 + 4\,\ln 2 \right) \nn \\
(\hat r^{\pm \pm}_{\msbar})_{32} & = & 
\pm \left( 1 - \ln 2 \right) \nn \\
(\hat r^{\pm \pm}_{\msbar})_{33} & = &
- {{2}\over {N_c}} - {{2\,\ln 2}\over {N_c}} + 4\,N_c \nn \\
(\hat r^{\pm \pm}_{\msbar})_{44} & = &
\mp \left( {{39}\over 8} - 4\,\ln 2\right) - {9\over {2\,N_c}} - 
 {{2\,\ln 2}\over {N_c}} + 4\,N_c \nn \\
(\hat r^{\pm \pm}_{\msbar})_{45} & = &
\mp \left( {{31}\over {96}} - {{2\,\ln 2}\over 3} \right)
+ {{17}\over {24\,N_c}} - {{5\,\ln 2}\over {6\,N_c}} + 
{{N_c}\over {16}} \\
(\hat r^{\pm \pm}_{\msbar})_{54} & = &
\pm \left( {{13}\over 2} + 32\,\ln 2 \right) + {{34}\over {N_c}} - 
 {{40\,\ln 2}\over {N_c}} - 7\,N_c \nn \\
(\hat r^{\pm \pm}_{\msbar})_{55} & = &
\mp \left( {{95}\over {24}} - {{28\,\ln 2}\over 3} \right) + 
{7\over {6\,N_c}} - {{26\,\ln 2}\over {3\,N_c}} \nn \\
(\hat r^{\pm \mp}_{\msbar})_{44} & = &
\pm {{17}\over 8} - {1\over {2\,N_c}} \nn \\
(\hat r^{\pm \mp}_{\msbar})_{45} & = &
\pm {3\over {32}} + {3\over {8\,N_c}} - {{N_c}\over {16}} \nn \\
(\hat r^{\pm \mp}_{\msbar})_{54} & = &
\pm {{21}\over 2} - {{22}\over {N_c}} + 7\,N_c \nn \\
(\hat r^{\pm \mp}_{\msbar})_{55} & = &
\mp {{13}\over 8} + {1\over {2\,N_c}} \nn
\eea

\section*{Acknowledgements}
V.L., G.M. and I.S. acknowledge the M.U.R.S.T.
and the INFN for partial support. L.S. acknowledges the support of 
Fondazione A. Della Riccia and of German Bundesministerium f\"{u}r 
Bildung and Forschung under contract 06 TM 874 and DFG Project 
Li 519/2-2.

\end{document}